\documentclass[12pt]{iopart}
\usepackage{amssymb}
\usepackage{epsfig}
\usepackage{subfigure}
\usepackage{graphicx}
\usepackage{textcomp}
\usepackage{cite}
\usepackage{float}
\usepackage{subfigure}
\usepackage{color}
\usepackage{ulem}
\expandafter\let\csname equation*\endcsname\relax
\expandafter\let\csname endequation*\endcsname\relax
\usepackage{amsmath}
\newcommand{\agt}{\mathrel{\raise.3ex\hbox{$>$\kern-.75em\lower1ex\hbox{$\sim$}}}}
\begin{document}

\title{Gradual Diffusive Capture: Slow Death by Many Mosquito Bites}

\author{S. Redner}
\address{Santa Fe Institute, 1399 Hyde Park Road, Santa Fe, New Mexico 87501, USA}
\address{Center for Polymer Studies and Department of Physics, Boston
  University, Boston, MA 02215, USA}

\author{O. B\'enichou} \address{Laboratoire de Physique Th\'eorique de la
  Mati\`ere Condens\'ee (UMR CNRS 7600), Universit\'e Pierre et Marie Curie,
  4 Place Jussieu, 75255 Paris Cedex France}

\begin{abstract}

  We study the dynamics of a single diffusing particle (a ``man'') with
  diffusivity $D_M$ that is attacked by another diffusing particle (a
  ``mosquito'') with fixed diffusivity $D_m$.  Each time the mosquito meets
  and bites the man, the diffusivity of the man is reduced by a fixed amount,
  while the diffusivity of the mosquito is unchanged.  The mosquito is also
  displaced by a small distance $\pm a$ with respect to the man after each
  encounter.  The man is defined as dead when $D_M$ reaches zero.  At the
  moment when the man dies, his probability distribution of displacements $x$
  is given by a Cauchy form, which asymptotically decays as $x^{-2}$, while the
  distribution of times $t$ when the man dies asymptotically decays as
  $t^{-3/2}$, which has the same form as the one-dimensional first-passage
  probability.

\end{abstract}
\pacs{02.50.Ey, 05.10.Gg, 05.40.Fb}
\maketitle

\section{Introduction}

One-dimensional diffusive capture processes have rich properties that have
inspired much research in the applied
probability~\cite{BG91,K92,B00,LS02,KT02} and statistical physics
communities~\cite{F84,RK84,BZK84,FG88,BMS13}.  One example of this genre is
the ``lamb-lion'' problem~\cite{KR96,RK99}, in which a lamb diffuses on the
infinite line in the presence of $N$ independently diffusing lions that
initially are all to one side of the lamb.  Whenever a lamb and lion meet,
the lamb is killed.  Asymptotically, the survival probability of the lamb due
to these $N$ lions decays as $S_N(t)\sim t^{-\beta_N}$, where $\beta_N$
depends non-trivially on the number of lions~\cite{KR96,RK99} and secondarily
on the diffusivities of the lamb and the lion, $D_\ell$ and $D_L$,
respectively.

When $D_\ell=D_L$, the qualitative dependence of $\beta_N$ is understood,
even though the exact value of $\beta_N$ is unknown for $N\geq 3$.  For
$N=1$, $\beta_1=\frac{1}{2}$, as this case can be mapped to the probability
that a single random walker---corresponding to the lamb-lion
separation---does not hit the origin by time $t$~\cite{F71,W94,R01}.  For
$N=2$, $\beta_2=\frac{3}{4}<2\beta_1$~\cite{KR96,RK99}; that is, two lions
are less effective in killing the lamb than what might be expected because of
their ostensible independence.  The resolution of this apparently
contradictory behavior is that whenever the lamb moves, the lions move in
unison with respect to the reference frame of the lamb.  This induced
correlation between the lions implies that the effective number of lions is
less than 2.  This three-body problem may be exactly solved by mapping it to
the diffusion of a single effective particle in an absorbing two-dimensional
wedge of opening angle $\frac{2\pi}{3}$ (and generally the opening angle is a
function of $D_\ell$ and $D_L$) whose solution recovers
$\beta_2=\frac{3}{4}$~\cite{FG88,KR96,RK99,R01}; moreover, $\beta_2$ is
readily computable for arbitrary $D_\ell$ and $D_L$~\cite{R01,CJ59}.

For $N\geq 3$, simulations give $\beta_3\approx 0.91342$~\cite{AJMKR03},
$\beta_4\approx 1.03$, and $\beta_{10}\approx 1.4$~\cite{BG91}, but an
analytical solution for $N\geq 3$ is not yet known.  While the $N$-lion
problem can be readily mapped onto the diffusion of an effective particle in
$N+1$ dimensions that is restricted to an absorbing wedge region defined by
$N$ constraint hyperplanes---the so-called Weyl chamber~\cite{G99}---the
solution to this simply-stated problem seems to be difficult to achieve.  A
simplification arises in the limit $N\to\infty$, however, where one can map
the problem to the diffusion of an effective particle in the presence of an
absorbing and approaching boundary whose position scales as
$\sqrt{t}$~\cite{B66,U80,S88,T92,I92,KR96b}.  This approach gives
$\beta_N\simeq \frac{1}{4}\ln N$ for $N<\infty$ and $S_\infty(t)\simeq
\exp(-\ln^2 t)$ for $N=\infty$~\cite{KR96,RK99}.

\begin{figure}[ht]
\centerline{\includegraphics[width=0.6\textwidth]{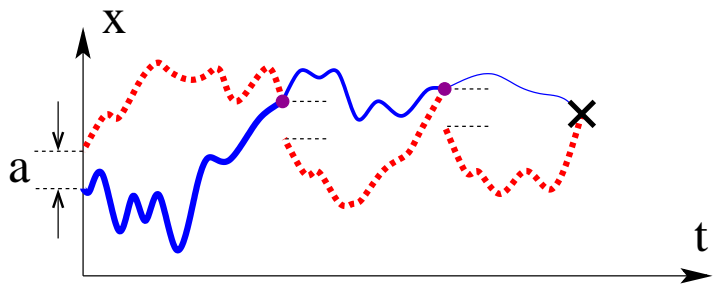}}
\caption{ Space-time evolution of a single mosquito (dashed) and man (solid).
  After each encounter, the mosquito is displaced by $\pm a$ with respect to
  the man, while the man becomes progressively more sluggish and eventually
  comes to rest.}
\label{model}
\end{figure}

Inspired by this lamb-lion system and also motivated by the unusual
phenomenology of locally activated random walks~\cite{BMRV12}, we introduce
the ``man-mosquitoes'' model (Fig.~\ref{model}), in which independent
``mosquitoes'', each with fixed diffusivity $D_m$, repeatedly bite the man
that has an evolving diffusivity $D_M$.  After each mosquito bite, the
diffusivity of the man is reduced by a fixed amount $\Delta D$, while the
diffusivity of each mosquito is unchanged.  To avoid spurious singularities,
the mosquito immediately moves equiprobably either a distance $+a$ or $-a$
from the man after each encounter.  Thus after each mosquito bite, the
encounter process begins anew, but with the diffusivity of the man reduced by
$\Delta D$.  For convenience, we define the initial diffusivity of the man
$D_M$ to be an integer multiple of $\Delta D$, that is, $D_M=\mathcal{N}(\Delta D)$.
Thus after being bitten $\mathcal{N}=D_M/\Delta D$ times, the man is at rest, which we
define as being ``dead''.  Our goal is to determine the distribution of times
and distances traveled after each encounter with the mosquitoes and the
position and time elapsed when the man dies.

Here we treat the simplest, but still non-trivial, case of a single mosquito.
Initially, the man is at $x_M=x_0$ and the mosquito is equiprobably at either
$x_m=x_0+ a$ or $x_0-a$, where $a$ can be viewed as the lattice spacing in
a discrete random-walk formulation.  The basic ingredient to understand how
the man moves is determined by the statistical properties of the first
encounter between the man and mosquito when they have arbitrary diffusivities
and start a distance $a$ apart.  The full problem involves convolving $\mathcal{N}$ of
these first encounters.  To solve for a single first encounter, we map the
one-dimensional motions of the two particles onto the motion of a single
effective particle in two dimensions, whose coordinates are the positions of
the man and the mosquito, and then apply the image method to solve this
effective problem (Sec.~\ref{single}).  In section~\ref{multiple}, we apply
these single-encounter results to determine the dynamics of multiple
encounters.  By this approach, we thereby determine the probability
distributions of displacements and lifetimes of the man at the moment when he
dies.  We provide a brief discussion and outline several extensions of this
work in Sec.~\ref{outlook}.

\section{Single Man-Mosquito Encounter}
\label{single}

\subsection{Equal Diffusivities}

As a preliminary, we review the first-passage properties of an isotropically
diffusing particle in two dimensions that is initially located at $(x,y)=(0,a)$,
diffuses in the half space $y>0$, and is absorbed when it hits any point $x$
along the locus $y=0$~\cite{F71,W94,R01}.  This absorption is characterized by: (i)
the first-passage probability, namely, the probability that the particle
first hits a point $(x,0)$ on the absorbing line at time $t$, and (ii) the
eventual hitting probability, the probability that particle is eventually
absorbed at $(x,0)$.

The probability density of the particle, $c(x,y,t)$, obeys the diffusion
equation, $\partial_t c=D\nabla^2 c$, subject to the absorbing boundary
condition $c(x,0,t)=0$ and the initial condition
$c(x,y,t\!=\!0)=\delta(x)\delta(y-a)$.  By the image method, the solution is
a sum of a Gaussian centered at $(0,a)$ and an anti-Gaussian centered at the
image point $(0,-a)$:
\begin{equation}
\label{c}
c(x,y,t)= \frac{1}{4\pi Dt}\left\{ e^{-[x^2+(y-a)^2]/4Dt} -
  e^{-[x^2+(y+a)^2]/4Dt} \right\}\,.
\end{equation}
From this expression, the first-passage probability to the point $(x,y=0)$ is
\begin{equation}
\label{j}
F(x,t)= D\frac{\partial c}{\partial y}\,\bigg|_{y=0} =\frac{a}{4\pi Dt^2}\,\,
e^{-(x^2+a^2)/4Dt}\,.
\end{equation}
Notice that while the flux, $-D\frac{\partial c}{\partial y}$, is in the $-y$
direction, the first-passage probability is positive.  Integrating this
first-passage probability over all time gives the eventually hitting
probability to a point $(x,0)$ on the $x$-axis.  Using the variable
substitution $u= (x^2+a^2)/4Dt$, this integration becomes elementary and the
result is the Cauchy, or Lorenztian, distribution:
\begin{equation}
\label{E}
E(x)= \int_0^\infty F(x,t)\,dt = \frac{1}{\pi}\, \frac{|a|}{x^2+a^2}~.
\end{equation}
The absolute value sign makes this result valid for $a>0$ and $a<0$.  Because
of this power-law decay, the mean-square position of the particle
when it hits the absorbing line is infinite.  Parenthetically, the
probability that the particle first hits {\it any} point on the $x$-axis at
time $t$ is simply the classic one-dimensional first-passage probability
\begin{equation}
\label{well known}
  f(t)=\int_{-\infty}^\infty\,\, \frac{|a|}{4\pi Dt^2}\,\,
  e^{-(x^2+a^2)/4Dt}\, dx = \frac{|a|}{\sqrt{4\pi Dt^3}}\,\, e^{-a^2/4Dt}\,,
\end{equation}
for which the mean first-hitting time is divergent.

\subsection{General Diffusivities}

\begin{figure}[ht]
\centerline{\includegraphics[width=0.4\textwidth]{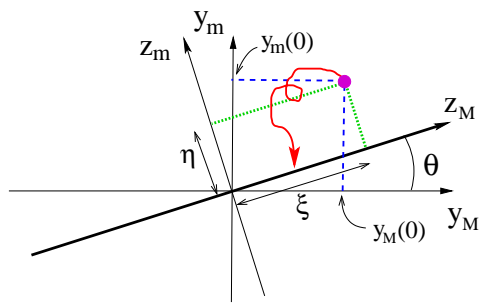}}
\caption{Mapping the diffusion of a man and mosquito on the line to diffusion
  in the half space $z_m>0$.  A first-passage trajectory is shown.}
\label{plane}
\end{figure}

We now study the case of a man with diffusivity $D_M$ and mosquito with
diffusivity $D_m$ that are initially at $x_M=x_0$ and $x_m=x_0+a$,
respectively; later we will average over the cases where initially
$x_m=x_0+a$ and $x_m=x_0-a$.  To determine the probability that the man and
mosquito first meet at position $x$ at time $t$ when starting from this
initial state, we map this two-particle system to an effective
single-particle problem in the two-dimensional coordinates $(x_M,x_m)$.  The
diffusion coefficient of the effective particle is anisotropic, since
typically $D_M\ne D_m$.  To simplify matters, we rescale the effective
particle coordinates to $y_M=x_M/\sqrt{D_M}$ and $y_m=x_m/\sqrt{D_m}$ so that
the effective particle diffuses isotropically with unit diffusivity.  The
condition that the mosquito meets the man, $x_M=x_m$, translates to
$y_M\sqrt{D_M}=y_m\sqrt{D_m}$.  Thus the effective particle starts at
$[x_0/\sqrt{D_M},(x_0\!+\!a)/\sqrt{D_m}]$, moves with unit diffusivity, and
eventually hits the line $y_M\sqrt{D_M}=y_m\sqrt{D_m}$ that is inclined at an
angle $\theta= \tan^{-1}(\sqrt{D_M/D_m})$ with respect to the horizontal
(Fig.~\ref{plane}).

To analyze this problem, it is convenient to introduce the rotated
coordinates $z_M,z_m$ defined by
\begin{align}
\begin{split}
z_M&= y_M\cos\theta + y_m\sin\theta = \frac{x_M}{\sqrt{D_M}}\cos\theta +\frac{x_m}{\sqrt{D_m}}\sin\theta\,,\\
z_m&= y_m\cos\theta - y_M\sin\theta = \frac{x_m}{\sqrt{D_m}}\cos\theta -\frac{x_M}{\sqrt{D_M}}\sin\theta\,,\\
\end{split}
\end{align}
because the mosquito meeting the man corresponds to the effective particle
hitting a point on the locus $z_m=0$.  In these rotated coordinates, the initial
condition becomes
\begin{align}
\begin{split}
z_M(0)&\equiv \xi 
= \frac{x_M(0)}{\sqrt{D_M}}\cos\theta+\frac{(x_M(0)+a)\sin\theta}{\sqrt{D_m}}
=\frac{x_M(0)}{\sqrt{D_M}\,\cos\theta}+\frac{a\sin\theta}{\sqrt{D_m}} \,,\\
z_m(0)&\equiv \eta 
= \frac{(x_M(0)+a)\cos\theta}{\sqrt{D_m}}-\frac{x_M(0)}{\sqrt{D_M}}\sin\theta
=  \frac{a\cos\theta}{\sqrt{D_m}}\,.
\end{split}
\end{align}
Following Eq.~\eqref{j}, the probability density of the effective particle in
the $z$-plane is again the sum of a Gaussian and an anti-Gaussian:
\begin{equation}
c(z_M,z_m,t)= \frac{1}{4\pi t}\left\{ e^{-[(z_M-\xi)^2+ (z_m-\eta)^2]/4t} -
 e^{-[(z_M-\xi)^2+ (z_m+\eta)^2]/4t} \right\}\,.
\end{equation}
From this expression, the first-passage probability to a point $z_M$ on the
$z_m=0$ axis is
\begin{equation}
\label{jz}
F(z_M,t)= \frac{|\eta|}{4\pi t^2}\,\,e^{-[(z_M-\xi)^2+\eta^2]/4t}\,,
\end{equation}
while the eventually hitting probability to this point, the time integral of
\eqref{jz}, is
\begin{equation}
\label{Ez}
E(z_M)= \frac{1}{\pi}\, \frac{|\eta|}{(z_M-\xi)^2+\eta^2}~,
\end{equation}
where the absolute value sign accounts for the possibility that the mosquito
could initially be on either side of the man.

To appreciate the meaning of these results, we need to transform the above
expressions for $E$ and $F$ to the original coordinates.  For the eventual
hitting probability $E$, the peak in this quantity occurs when $z_M=\xi$.
From Fig.~\ref{plane}, this condition translates to
\begin{equation}
x_M=\sqrt{D_M}\,\cos\theta\, z_M = \sqrt{D_M}\,\cos\theta\, 
\bigg[\frac{x_M(0)}{\sqrt{D_M}}\cos\theta +
\frac{x_M(0)+a}{\sqrt{D_m}}\sin\theta\bigg]\,,
\end{equation}
which we recast, after some simple algebra, as
\begin{equation}
x\equiv x_M-x_M(0)=a\sin^2\theta = a\,\,\frac{D_M}{D_M+D_m}~.
\end{equation}
As expected, if the mosquito diffuses quickly, the man will have barely moved
when he is bitten.  Conversely, if the mosquito diffuses slowly, the man will
be close to the initial position of the mosquito when the bite occurs.

We now use $E(z)dz=E(x)dx$ to transform \eqref{Ez} to the original $x_M,x_m$
coordinates to give:
\begin{subequations}
\label{Es}
\begin{equation}
\label{E-theta}
E(x) = \frac{1}{\pi}\,\,\frac{|a|\sin\theta\cos\theta}{\big(x- a
  \sin^2\theta\big)^2+ \big(a\sin\theta\cos\theta\big)^2}
\end{equation}
for the probability that the man has moved a distance $x$ when he is first
bitten by the mosquito.  Using $\tan\theta=\sqrt{D_M/D_m}$, we can
equivalently rewrite $E(x)$ in terms of the diffusivities only:
\begin{equation}
\label{E-D}
E(x) = \frac{|a|}{\pi}\,\,\frac{\sqrt{D_MD_m}}{(D_M+D_m)}
\left\{\left[x-  a\,\frac{D_M}{D_M+D_m}\right]^2
+ \left[a\,\frac{\sqrt{D_MD_m}}{(D_M+D_m)}\right]^2\right\}^{-1}~,
\end{equation}
\end{subequations}
This expression holds for $a>0$ and $a<0$, and we should average over these
two cases to account for the possibilities that the mosquito is initially to
the right or to the left of the man.  Thus the probability distribution of
displacements after a single encounter with the mosquito again has the
long-range Cauchy form first given in \eqref{E}.  Similarly, Eq.~\eqref{jz}
can be transformed back to the original variables, to give the probability
$F(x,t)$ that the man has moved a distance $x$ when he meets the mosquito at
time $t$:
\begin{align}
\label{Fxt}
F(x,t)=\frac{|a|}{4\pi t^2 \sqrt{D_M
 D_m}}\,\,e^{-\frac{1}{4t}\left\{\left(x\sqrt{\frac{D_M+D_m}{D_MD_m}}
-a\sqrt{\frac{D_M}{D_m(D_M+D_m)}}\right)^2+\left(\frac{\scriptstyle a}{\sqrt{D_M+D_m}}\right)^2\right\}}~.
\end{align}
Integrating this last expression over all positions, the distribution of
times when the man first encounters the mosquito is
\begin{align}
\label{ft}
f(t)=\frac{|a|}{\sqrt{4\pi(D_M\!+\!D_m)\,t^3}}\,\,e^{-a^2/[4(D_M+D_m)t]}\,.
\end{align}
This result is well known and is easily obtained from Eq.~\eqref{well known}
because if one is concerned only with the meeting time, one only needs the
separation between the man and mosquito, and this variable diffuses with
diffusion coefficient $D_M+D_m$.

\section{Multiple Encounters}
\label{multiple}

We now compute the probability that the man has moved a distance $x$ at the
moment when he dies, which occurs when the man has been bitten $\mathcal{N}$ times.  As
a preliminary, the Fourier transform of the Cauchy step-length distribution
\eqref{Es} at the $n^{\rm th}$ encounter with the mosquito (with $n\leq \mathcal{N}$)
is:
\begin{align}
\label{Ek}
E_n(k)\equiv\int E_n(x)\,e^{ikx}dx &=\frac{1}{\pi}\int
\frac{B_{n-1}}{(x-A_{n-1})^2+B_{n-1}^2}\,\,\, e^{ikx}\, dx
 = e^{ikA_{n-1}-|k|B_{n-1}}\,.
\end{align}
Here 
\begin{equation*}
A_n=a\epsilon_n\,\,\frac{D_M(n)}{D_M(n)+D_m(n)}\,,\qquad\qquad
B_n=a\,\,\frac{\sqrt{D_M(n)D_m(n)}}{D_M(n)+D_m(n)}\,,
\end{equation*}
as given in \eqref{E-D}.  Here $\epsilon_n=\pm1$ is a random variable that
assumes the values $+1$ or $-1$, respectively, if the mosquito is displaced
by $+a$ or $-a$ after the $n^{\rm th}$ meeting with the man; these two events
occur with probability $1/2$.  Additionally, all relevant variables are now
indexed by $n$, the number of man-mosquito encounters.  For a given
realization of $\epsilon_0, \epsilon_1,\cdots,\epsilon_\mathcal{N}$, the
probability distribution of displacements of the man at the $\mathcal{N}^{\rm
  th}$ encounter with the mosquito equals the product of the Fourier
transforms of single-encounter displacement distributions:
\begin{equation}
\label{MEk}
\mathcal{E}_\mathcal{N}(k)\equiv \prod_{n=1}^{\mathcal{N}} E_n(k) = \exp\Big[\sum_{n=0}^{\mathcal{N}-1}(ikA_n-|k|B_n)\Big]\,.
\end{equation}
Inverting this expression (see Eq.~\eqref{Ek}), the probability distribution
at the moment when the man and mosquito have met $\mathcal{N}$ times, and for a given
realization $\epsilon_0, \epsilon_1,\cdots,\epsilon_\mathcal{N}$, is
\begin{equation}
\label{PN}
\mathcal{E}_\mathcal{N}(x)=\frac{1}{\pi} \frac{\sum_n B_n}{\big(x-\sum_{n=0}^{\mathcal{N}-1} A_n\big)^2+\big(\sum_{n=0}^{\mathcal{N}-1} B_n\big)^2}\,.
\end{equation}

For the specific case where initially $D_M(0)=D_m(0)=1$ and the diffusivity
of the man decreases by $1/\mathcal{N}$ each time the mosquito bites the man, then
$D_M(n)=\big(1-\frac{n}{\mathcal{N}}\big)$ and the man dies after $\mathcal{N}$ bites.  For this
choice
\begin{equation*}
A_n= a\epsilon_n\,\,\frac{1-{n}/{\mathcal{N}}}{2-{n}/{\mathcal{N}}}\equiv a\epsilon_n\, f(n/\mathcal{N})\,,\qquad\qquad
B_n= |a|\,\, \frac{\sqrt{1-{n}/{\mathcal{N}}}}{2-{n}/{\mathcal{N}}}\equiv |a|\, g(n/\mathcal{N})\,.
\end{equation*}
To estimate the average of the probability distribution \eqref{PN} over
all realizations of $\epsilon_0,\epsilon_1,\epsilon_2,\cdots,\epsilon_\mathcal{N}$, we need
$\sum_n A_n$ and $\sum_n B_n$.  Because $f(z)$ and $g(z)$ are slowly varying
functions of $z$, the leading behaviors of these sums are $\sum_n A_n\simeq
\mathcal{N}^{1/2}\, a\,\alpha$, and $\sum_n B_n\simeq \mathcal{N}\,a\,\beta$, where
$\alpha,\beta$ are of the order of one.  Thus the probability distribution of
displacements of the man when he dies is again the Cauchy distribution
\begin{equation}
\label{PN-asym}
\mathcal{E}_\mathcal{N}(x)=\frac{1}{\pi}\, \frac{\mathcal{N}\,|a|\,\beta}{(x\!-\!
  \mathcal{N}^{1/2}\,a\,\alpha)^2 +(\mathcal{N}\, a\, \beta)^2}
\sim \frac{1}{\pi} \,\frac{\mathcal{N}\,|a|\,\beta}{x^2+(\mathcal{N}\, a\, \beta)^2}~.
\end{equation}
Thus, while the typical distance that the man moves before he dies is of the
order of $\mathcal{N}$, the long tail of this distribution leads to the average
distance traveled being infinite.  In this result, the displacement of the
mosquito by $\pm a$ after each bite contributes only to the subdominant term
$\mathcal{N}^{1/2}a\alpha$ in $\mathcal{E}_\mathcal{N}(x)$ and therefore is asymptotically
irrelevant.

Following this same approach, we may also compute the probability $f_\mathcal{N}(t)$
that the man dies at time $t$ (which occurs after $\mathcal{N}$ mosquito bites).  This
probability is the convolution of the single-encounter first-passage
probabilities~\eqref{ft}.  For this convolution, we first need the
Laplace transform of Eq.~\eqref{ft}.  This is given by
\begin{align}
f(s)=\int_0^\infty f(t)\,e^{-st}\, dt &=
\int_0^\infty
\frac{|a|}{\sqrt{4\pi(D_M\!+\!D_m)\,t^3}}\,\,e^{-a^2/[4(D_M+D_m)t]}\,e^{-st} dt\,,\nonumber \\
&=\phantom{\frac{1}{2}}e^{-|a|\sqrt{s/(D_M+D_m)}}~.
\end{align}
For $\mathcal{N}$ encounters with the mosquito, the Laplace transform is given by
\begin{align}
f_\mathcal{N}(s)&=\prod_{n=0}^{\mathcal{N}-1} e^{-|a|\sqrt{s/(2-n/\mathcal{N})}}~,
\end{align}
where we have used $D_m=1$ and $D_M=\big(1-\frac{n}{\mathcal{N}}\big)$.
We now write the product as the exponential of the sum and replace the sum by
an integral.  The resulting integral is elementary and the result is
\begin{equation}
f_\mathcal{N}(s)= e^{-w\sqrt{s}}\,,
\end{equation}
where $w=2(\sqrt{2}-1)|a|\mathcal{N}$.  Inverting this Laplace transform gives
the distribution of times when the man dies, which happens after $\mathcal{N}$
encounters with the mosquito:
\begin{equation}
f_\mathcal{N}(t)=\frac{w}{\sqrt{4\pi t^3}}\,\, e^{-w^2/4t}\,.
\end{equation}
For large times, this first-passage probability at the moment of death decays
as $\mathcal{N}/t^{3/2}$.  The mean time to die is infinite, as is the case for a
single man-mosquito encounter.  However, as $\mathcal{N}$ increases, corresponding to
each mosquito bite being less damaging, the amplitude of the asymptotic
distribution increases linearly with $\mathcal{N}$.

\section{Discussion}
\label{outlook}

We introduced a diffusive capture process in which each interaction between a
prey and a predator causes the diffusivity of the prey to decrease by a fixed
amount.  After a finite number of interactions of this type, the prey
eventually comes to rest, which we define as its death.  As in the
conventionally-studied case where the predator kills the prey when they first meet,
basic questions about this capture process are the survival probability of
the prey and the properties of its motion during the chase.

In this work, we investigated the simple case of a single predator, for which
the above long-time properties of the prey can be obtained from classic
first-passage concepts.  Specifically, we determined the distribution of
displacements when the prey dies.  Between each encounter with the mosquito,
the displacement distribution is given by the long-ranged Cauchy
distribution.  Thus the displacement distribution when the man dies---which
is a convolution of single-encounter distributions---also has a Cauchy form
that is given by Eq.~\eqref{PN-asym}.  Thus even though the prey is
continuously hunted by the predator and its health---quantified by its
diffusivity---is progressively declining, the prey can, on average, endure to
wander over a wide range.

There are a number of natural directions for further work.  The most
immediate is the case of $N\geq 2$ mosquitoes, where the man has to escape
a cloud of mosquitoes rather than just a single mosquito.  The special case
of $N=2$ mosquitoes should be amenable to an asymptotic solution because the
dynamics of this three-particle system after each mosquito bite is, in
principle, known.  When the mosquitoes and the man initially have the same
diffusivities, the probability that the first bite occurs at time $t$ or
greater decays as $t^{-3/4}$, for the initial ordering $Mmm$
($M=\mathrm{man}$, $m=\mathrm{mosquito}$), and as $t^{-3/2}$, for the initial
ordering $mMm$.  (Note that the relative ordering of the man and the mosquito
is important only in one dimension.)~ After many mosquito bites, the
corresponding exponents for these same orderings are~\cite{R01}:
\begin{align*}
&\beta_{\rm Mmm}= 
\bigg[2-\frac{2}{\pi}\cos^{-1}\bigg(\frac{D_M}{1+D_M}\bigg)\bigg]^{-1}\sim
1 -\frac{D_M}{\pi}  \qquad D_M\to 0\,,\\
&\beta_{\rm mMm}=
\pi\left[{2\cos^{-1}\bigg(\frac{D_M}{1+D_M}\bigg)}\right]^{-1}\hskip 0.6cm \sim
1+D_M\qquad D_M\to 0\,.
\end{align*}
Thus as the man is repeatedly bitten and becomes more sluggish, his survival
probability decays with an exponent that approaches one as $D_M\to 0$,
independent of the relative positions of the man and the mosquitoes.  The
effect of this exponent approaching 1 is that that the average lifetime of
the man is infinite, even if he is initially surrounded by the mosquitoes, a
configuration that would lead to a finite lifetime when a single mosquito
bite is fatal.

For larger $N$, the basic question is to determine the dependence of the
lifetime of the man as a function of $N$ and the initial diffusivities of the
man and the mosquitoes.  In general, when a single mosquito bite is fatal,
the mean lifetime of the man is finite (the only exceptions are the cases of
$N=1$, as well as $N=2$ and $N=3$, with all the mosquitoes initially to one
side of the man).  It will be worthwhile to determine the dependence of the
lifetime on $N$ and on the virulence of the mosquitoes, as quantified by
$\Delta D$, the decrease in the diffusivity of the man after each bite.  What
is an open question is the nature of the dependence of the man's lifetime on
basic parameters in two dimensions or greater.

The approach developed here can, in principle, be extended to compute the
joint distribution that the man has traveled a distance $x$ and has lived for
a time $t$ when he dies.  Starting with the joint probability \eqref{Fxt}
that the man has traveled a distance $x$ over a time $t$ when an encounter
occurs, the corresponding distribution after $\mathcal{N}$ encounters with the mosquito
is the convolution of these single-encounter joint probabilities.  The
calculation of these convolutions is conceptually straightforward by making
use of the Laplace-Fourier transform technique.  However, the inversion of
this transforms does not appear to be tractable.

\medskip We thank Raphael Voituriez for helpful discussions.  We also thank NSF
Grant No.\ DMR-1205797 (SR) and ERC starting Grant No.\ FPTOpt-277998 (OB)
for partial support of this research.

\end{document}